\documentclass[preprint,aps,floatfix,nofootinbib,superscriptaddress]{revtex4-1}

\usepackage{amsmath, amssymb, amsfonts, amsthm, latexsym, epsfig, mathrsfs, xcolor, bbm}

\usepackage[inline]{enumitem}
\usepackage[utf8]{inputenc}
\usepackage{setspace}
\usepackage[marginal, multiple]{footmisc}

\usepackage[utf8]{inputenc}

\usepackage[colorlinks, allcolors=blue!70!black, linktocpage]{hyperref}

\setlist[enumerate]{noitemsep, label=(\arabic*), ref=(\arabic*)}

\numberwithin{equation}{section}

\usepackage{cleveref}

\crefname{equation}{Eq.}{Eqs.}
\creflabelformat{equation}{#2#1#3}

\crefname{section}{Sec.}{Sec.}
\crefname{appendix}{Appendix}{Appendices}
\crefname{figure}{Fig.}{Figs.}

\crefname{definition}{Def.}{Defs.}
\crefname{prop}{Prop.}{Props.}
\crefname{lemma}{Lemma}{Lemmas}
\crefname{corollary}{Cor.}{Cors.}
\crefname{thm}{Theorem}{Theorems}
\crefname{remark}{Remark}{Remarks}

\makeatletter
\def\p@subsection{}
\def\p@subsubsection{}
\makeatother

\let\OLDfootnote\footnote
\renewcommand\footnote[1]{%
	\setlength{\footnotesep}{0.55\baselineskip}%
	{\footnotesize \OLDfootnote{#1}}%
}

\let\OLDthebibliography\thebibliography
\renewcommand\thebibliography[1]{%
	\setstretch{1.079} 
	\OLDthebibliography{#1}%
	\small %
	\setlength{\itemsep}{0.2\baselineskip} 
}

\newcommand{\M}{\mathcal{M}}
\newcommand{\MM}{\widetilde{\mathcal{M}}}
\newcommand{\Lg}{\mathcal{L}}
\newcommand{\OO}{\mathcal{O}}
\newcommand{\rmax}{r_{\text{max}}}

\newcommand{\x}{\mathbf{x}}
\newcommand{\is}{\int_{\Sigma} d^{d-1}x \sqrt{h(x)}}

\newcommand{\bra}[1]{\langle #1|}
\newcommand{\ket}[1]{|#1\rangle}

\newcommand{\braket}[2]{\langle #1|#2\rangle}

\def\hhmm{\number\hh:\ifnum\mm<10{}0\fi\number\mm}

\def\be{\begin{equation}}
\def\ee{\end{equation}}

\begin{document}
	
	\setstretch{1.2}
	
	\title{Is volume the holographic dual of fidelity susceptibility$?$}
	\date{November 20, 2018}
	\author{Mudassir Moosa} \email{mudassir.moosa@cornell.edu}\affiliation{Center for Theoretical Physics and Department of Physics \\ University of California, Berkeley, CA 94720, USA \\ Lawrence Berkeley National Laboratory, Berkeley, CA 94720, USA}\affiliation{Department of Physics, Cornell University, Ithaca, NY, 14853, USA} \author{Ibrahim Shehzad}\email{is354@cornell.edu}
	\affiliation{Department of Physics, Cornell University, Ithaca, NY, 14853, USA}

	\begin{abstract}
		It was proposed by Miyaji et al. that the fidelity susceptibility of a state of a conformal field theory under a marginal deformation is holographically dual to the volume of a maximal time slice in the dual Anti de Sitter spacetime. We study this proposal by analyzing the leading and subleading divergences in these two quantities in two specific scenarios. We find that although the structure of the divergences in these two quantities is similar, their numerical coefficients are inconsistent with an exact relationship between these two quantities.
		
	\end{abstract}

	\maketitle
	\tableofcontents
	
	\section{Introduction}
	
	The idea that the entanglement entropy of a spatial subregion in a conformal field theory (henceforth CFT) is related to the area of a particular spacelike surface in the dual asymptotically locally Anti de Sitter (henceforth AdS) spacetime \cite{RT, HRT, Proof-RT, Proof-HRT} has provided various insights into the nature of the AdS-CFT correspondence \cite{AdSCFT}. Not only does it provide a useful tool for computing entanglement entropy in the CFT, it also indicates that there is a deep connection between geometrical quantities (in this case area) in the bulk and information theoretic quantities (in this case entanglement) defined on the boundary. Motivated by this connection, there have been various recent proposals for an information theoretic quantity dual to the volume of a spatial slice in the bulk. One such conjecture relates the volume of a bulk spatial slice to the computational complexity \cite{C1,C2,ad1,MA} of a boundary state \cite{CequalsV1, CequalsV2}. A second conjecture relates it to the Kahler form induced by the overlap of certain nearby states \cite{Belin:2018fxe}. A third conjecture relates this volume to the change in the boundary state under a marginal deformation, measured by  \textit{fidelity susceptibility} (defined in Eq.~\eqref{eq-def-fid-pure}), as presented in \cite{MIyaji:2015mia}. Our goal in this paper is to investigate the latter.
	\newline

	
	The fidelity susceptibility, also called the \textit{Bures metric} or the \textit{quantum information metric}\footnote{An alternative metric, known as the quantum Fisher information metric, has been studied in \cite{Lashkari, Lashkari2,fisher}.}, is an important information theoretic quantity in and of itself. For instance, it is used to study quantum phase transitions and chaos in quantum systems \cite{Bak:2017rpp, Miyaji2} (also see \cite{Miyaji2} for a more detailed list of references). For the purposes of the discussion here, this metric is a measure of the distance between two quantum states related by a perturbation. The fidelity susceptibility, denoted by $G_{\lambda\lambda}$, for two pure states that differ by a perturbation labeled by $\lambda$, is defined via the following relation
	
	\begin{equation}
	|\braket{\psi(0)}{\psi({\lambda})}| \equiv 1 - G_{\lambda \lambda} \, \lambda^{2} + {O}(\lambda^{3}) \, . \label{eq-def-fid-pure}
	\end{equation}
	
	For the sake of completeness, let us point out that fidelity susceptibility can also be defined more generally for mixed states \cite{Uhl} but that will not concern us here. 
	\newline

	%

	
	Before proceeding further, let us include a brief review of the conjecture presented in \cite{MIyaji:2015mia} and in doing so lay out our notation. We assume that we have a CFT on a $d$-dimensional spacetime of the form $\M = R\times \Sigma$,  where $\Sigma$ is some codimension-1 spacelike hypersurface, and a theory that is obtained by a deformation of this CFT by a marginal operator, $\OO$. That is, the Lagrangian density of the deformed theory, $\Lg_{\lambda}$, is related to the Lagrangian density of the CFT, $\Lg_{0}$, by
	\begin{equation}
	\Lg_{\lambda} \, = \, \Lg_{0} +  \lambda \OO \, ,
	\end{equation}
	where the conformal dimension, $\Delta$, of $\OO$ is equal to the spacetime dimension $d$	and $\lambda$ is an infinitesimal coupling constant. As alluded to earlier, one calculates the fidelity susceptibility between the vacuum states of these two theories using Eq.~\eqref{eq-def-fid-pure}. For the states of a CFT with a classical holographic dual, it was conjectured in \cite{MIyaji:2015mia} that the fidelity susceptibility with respect to an infinitesimal marginal deformation is holographically dual to the volume of a bulk codimension-$1$ spacelike slice with maximal volume in the asymptotically locally AdS spacetime. More precisely, the 
	relation proposed in \cite{MIyaji:2015mia} was
	
	\be \label{eq-proposal}
	G_{\lambda \lambda} = n_{d} \frac{V_{\text{max}}}{\ell_{\mbox{\tiny{AdS}}}^{d}}\,,
	\ee
	
	where $n_{d}$ is $O(1)$ constant, $\ell_{\mbox{\tiny{AdS}}}$ denotes the AdS radius, and $V_{\text{max}}$ denotes the volume of a boundary-anchored codimension-$1$ spacelike slice with maximal volume in the asymptotically locally AdS spacetime. The conjecture Eq.~\eqref{eq-proposal} has been studied and applied in various contexts in \cite{Bak:2017rpp,Miyaji2,Trivella:2016brw,Momeni:2017kbs,Flory:2017ftd,Momeni:2017mmc}. 
	As an example where this conjecture works, 
	consider the vacuum state of a CFT on $R^{d}$. As derived in Appendix~(\ref{app-der}), the fidelity susceptibility of the vacuum state under a marginal deformation is given by 
	\begin{align}
	G^{(\text{vac})}_{\lambda\lambda} [R^{d}] \, = \, \int_{\epsilon}^{\infty} d\tau \int_{-\infty}^{-\epsilon} d\tau^{\prime} \int_{R^{d-1}} d^{d-1}x \int_{R^{d-1}} d^{d-1} x^{\prime}  \, \langle \OO(\tau,x) \OO(\tau^{\prime}, x^{\prime})\rangle_{R^{d}} \, , \label{eq-vac-fid-rd}
	\end{align}
	where $\langle ... \rangle_{R^{d}}$ is the vacuum expectation value in the original CFT on $R^{d}$. The integrals in Eq.~\eqref{eq-vac-fid-rd} diverge when we take $\epsilon \to 0$ limit. This is because the two point function diverges when the two operators are coincident. Therefore, $\epsilon$ plays the role of an ultraviolet (henceforth UV) cutoff. Recall that in a Euclidean spacetime with coordinates $(\tau,x)$, the two point function for a primary operator of conformal dimension $\Delta$ is given by
	
	\be \label{2pt}
	\langle \OO(\tau,x) \OO(\tau^{\prime}, x^{\prime})\rangle_{R^{d}} = \frac{1}{((\tau - \tau^{\prime})^{2} + (x - x^{\prime})^{2})^{\Delta}} \,.
	\ee
	
	One plugs Eq.~\eqref{2pt} into Eq.~\eqref{eq-vac-fid-rd} for a marginal operator, that is, for $\Delta=d$, and evaluates these integrals to get
	
	\be \label{cft-calc}
	G^{(\text{vac})}_{\lambda \lambda} [R^{d}] = N_{d} V_{d-1} \epsilon^{1-d}\,,
	\ee
	where $V_{d-1}$ is the infinite volume of $R^{d-1}$ and $N_{d}$ is an $O(1)$ constant whose explicit value will not concern us here. Recall also that the bulk dual of the vacuum state of the boundary CFT is Poincaré AdS spacetime with the metric
	\begin{align}
	ds^{2} \, = \, \frac{\ell_{\mbox{\tiny{AdS}}}^{2}}{z^{2}} \, \Big(dz^{2} - dt^{2} + d\mathbf{x}_{d-1}^{2}\Big) \, .
	\end{align}
	where z is a radial coordinate in the bulk and $(t,x)$ are defined as usual. Since this is a stationary spacetime, a constant time slice has vanishing extrinsic curvature and is therefore a maximal volume slice. The volume of this maximal volume slice in this geometry is then given by
	
	\be \label{eq-vol-poin}
	\frac{V_{\text{max}}}{\ell_{\mbox{\tiny{AdS}}}^{d}} = \int_{R^{d-1}} dx^{d-1} \, \int_{\delta}^{\infty} \frac{dz}{z^{d}} = \frac{V_{d-1}}{d-1} \, \delta^{1-d} \, ,
	\ee
	where a cutoff is introduced at $z=\delta$ near the AdS boundary. Since the field-theoretic cutoff and holographic cutoff are related by $\epsilon \sim \delta$ (discussed in Appendix \ref{app-cutoff}), one finds that the fidelity susceptibility in Eq.~\eqref{cft-calc} and the volume in Eq.~\eqref{eq-vol-poin} are indeed related according to the conjecture in Eq.~\eqref{eq-proposal}. Notice that the value of the proportionality constant, $n_{d}$, in Eq.~\eqref{eq-proposal} is undetermined in the example above since it depends on the exact relationship between the cutoffs $\epsilon$ and $\delta$. However, note that the coefficients of logarithmic divergences are independent of the choice of the cutoff. Hence, if there are logarithmic divergences in the fidelity susceptibility and in the volume of the bulk slice, then we can compare their coefficients to determine the value of $n_{d}$. This will play an important role in our analysis.
	\newline
	
	In our calculations, we assume that the $O(1)$ constant, $n_{d}$, in Eq.~\eqref{eq-proposal} is not cutoff dependent. We point out that the conjecture cited in Eq.~\eqref{eq-proposal} may still be valid if a cutoff dependence is allowed.
	\newline
	
	In the rest of this paper, we study the conjecture cited in Eq.~\eqref{eq-proposal} in two specific cases.   In Sec.~(\ref{sec-hyperbolic}), we study the UV divergences in the fidelity susceptibility  of a thermofield double (henceforth TFD) state of two copies of a CFT on $R\times H^{d-1}$ and compare it to the divergences in the volume of a maximal slice in the dual AdS black hole spacetime. In Sec.~(\ref{sec-sphere}), we repeat this analysis for the vacuum state of a CFT on $R\times S^{d-1}$ and the divergences in the volume of a maximal slice in the dual global AdS spacetime. We end with our conclusions in Sec.~(\ref{sec-conc}). A brief review of the definition of fidelity susceptibility is included in Appendix~(\ref{app-der}) while in Appendix~(\ref{app-cutoff}), we derive the relationships we use between the bulk and CFT cutoffs.

	\section{Fidelity susceptibility on $R\times H^{d-1}$} \label{sec-hyperbolic}
	
	In this section, our goal is to consider a TFD state of two copies of a CFT on $\M_{1} \, = \, R \times H^{d-1}$ and study its fidelity susceptibility under a deformation by a marginal operator, $\OO$. As derived in Appendix (\ref{app-der}), this fidelity susceptibility is related to the two-point function of the CFT on $\MM_{1} = S^{1}\times H^{d-1}$, where the period of $S^{1}$ is equal to the inverse temperature, $\beta$, of the TFD state. For simplicity, we assume\footnote{Although the generalization to arbitrary radii is straightforward, our choice suffices for the example we wish to present.} that the radii of the $S^{1}$ and that of the $H^{d-1}$ are both equal to 1 and therefore the inverse temperature satisfies $\beta = 2\pi$.  In this case, the fidelity susceptibility of the TFD state, according to Eq.~\eqref{eq-ther-fid-int}, becomes
	\begin{align}
	G_{\lambda\lambda}^{(\text{TFD})} \, [ \M_{1}] \, = \, &\int_{\epsilon}^{\pi  - \epsilon} d\theta_{1} \int_{-\pi + \epsilon}^{-\epsilon} d\theta_{2} \, \int_{0}^{\infty} dr_{1} \frac{1}{r_{1}^{d-1}} \, \int_{0}^{\infty} dr_{2} \frac{1}{r_{2}^{d-1}} \nonumber\\ \times& \,  \int d^{d-2}\x_{1} \int  d^{d-2}\x_{2}  \,\,\, \Big\langle \OO(\theta_{1},u_{1},\Omega_{1}) \OO(\theta_{2}, u_{2},\Omega_{2})\Big\rangle_{\MM_{1}} \, , \label{eq-fid-hyperbolic}
	\end{align}
	where $-\pi \le \theta \le \theta$ is the coordinate on $S^{1}$, whereas $r \ge 0$ and $x^{i}$ for $i = 1,2,...,d-2$ are the coordinates on $H^{d-1}$. The boundary of $H^{d-1}$ is at $r=0$ in these coordinates. In these coordinates, the metric of $\MM_{1} = S^{1}\times H^{d-1}$ is
	\begin{align}
	ds^{2} \, = \, d\theta^{2} + \, \frac{dr^{2} + d\x^{2}_{d-2}}{r^{2}} \ . \label{eq-met}
	\end{align}

	Note that we can write the metric of $\MM_{1}$ as
	\begin{align}
	ds^{2} \, = \, \frac{1}{r^{2}} \Big(dr^{2} + r^{2}d\theta^{2} + d\x^{2}_{d-2}\Big) \ .
	\end{align}
	Therefore, $\MM_{1}$ is conformally equivalent to $R^{d}$. As a result, we can write the two-point function on $\MM_{1}$ using the two-point function on $R^{d}$. Recall that the operator $\OO$ is exactly marginal with conformal dimension $\Delta = d$. Therefore, we get
	\begin{align}
	\Big\langle \OO(\theta_{1},r_{1},\x_{1})\OO(\theta_{2},r_{2},\x_{2})\Big\rangle_{\MM_{1}} \, = \, \frac{r_{1}^{d} \, r_{2}^{d} }{ \big[ r_{1}^{2} + r_{2}^{2} - 2 r_{1} r_{2} \cos(\theta_{1}-\theta_{2})  + (x_{1}-x_{2})^{2}\big]^{d} } \, . \label{eq-corr}
	\end{align} 
	
	
	Now we perform the integrals in Eq.~\eqref{eq-fid-hyperbolic}. Performing the integrals over $\x_{1}$ and $\x_{2}$ yields
	\begin{align}
	G_{\lambda\lambda}^{(\text{TFD})} \, [ \M_{1}] \, = \, N^{'}_{d} \, V_{R^{d-2}} \, \int_{\epsilon}^{\pi-\epsilon}& d\theta_{1} \int_{-\pi+\epsilon}^{-\epsilon} d\theta_{2} \, \int_{0}^{\infty} dr_{1} \, \int_{0}^{\infty} dr_{2} \, \, \frac{r_{1} \, r_{2} }{ \big[  r_{1}^{2} + r_{2}^{2} - 2 r_{1} r_{2} \cos(\theta_{1}-\theta_{2}) \big]^{\frac{d+2}{2}} } \, .  \label{eq-fid-ra}
	\end{align}
	where $N^{'}_{d}$ is an $O(1)$ constant. We now define $y_{\pm} \, = \, r_{1} \pm r_{2}$ and $\phi_{\pm} = (\theta_{1} \mp \theta_{2})/2\, $. With this change of variables, Eq.~\eqref{eq-fid-ra} becomes
	\begin{align}
	G_{\lambda\lambda}^{(\text{TFD})} \, [ \M_{1}] \, = \, \frac{N^{'}_{d}}{2} \, V_{R^{d-2}} \, \int_{\epsilon}^{\pi/2}& d\phi_{+} \int_{-\phi_{+}+\epsilon}^{\phi_{+}-\epsilon} d\phi_{-} \, \int_{0}^{\infty} dy_{+} \, \int_{-y_{+}}^{y_{+}} dy_{-} \, \, \frac{y^{2}_{+} - y^{2}_{-} }{ \left[ y_{+}^{2} \sin^{2}\phi_{+}   + y_{-}^{2} \cos^{2}\phi_{+} \right]^{\frac{d+2}{2}} } \, .\label{eq-fid-y}
	\end{align}
	
	Until this point, our analysis was completely general. From here on, we specialize to $d=5 \,$\footnote{As will become clear shortly, our argument requires a logarithmic divergence and at least two power law divergences in the fidelity susceptibility and in the volume of the bulk slice. The reason we are choosing $d=5$ is that this is smallest number of dimension in which we get two power law divergences and a logarithmic divergence.}.  With this value of $d$, we perform the integrals over $\phi_{-}$ and over $y_{-}$ to get
	\begin{align}
	G_{\lambda\lambda}^{(\text{TFD})} \, [ \M_{1}] \, = \, \frac{2 N^{'}_{5}}{15} \, V_{H^{4}} \, \int_{\epsilon}^{\pi/2}& d\phi_{+} \, \big( \phi_{+}-\epsilon \big) \,\,\, \frac{9-\cos(2\phi_{+})}{\sin^{6}\phi_{+}}    \, , \label{eq-fid-th}
	\end{align}
	where we have identified the volume of $H^{4}$ by 
	\begin{align}
	V_{H^{4}} \, = \, V_{R^{3}} \, \int_{0}^{\infty} dy_{+} \frac{1}{y_{+}^{4}} \, .
	\end{align}
	
	
	We now extract the divergences in the integral in Eq.~\eqref{eq-fid-th} to get
	\begin{align} \label{fid-div-sec2}
	G_{\lambda\lambda}^{(\text{TFD})} \, [ \M_{1}] \, = \, \frac{4 N^{'}_{5}}{75} \, V_{H^{4}} \, \left( \frac{1}{ \epsilon^{4}} + \frac{25}{ 6 \epsilon^{2}} - 14 \log\epsilon  \right) \, +(\cdots) \, ,
	\end{align}
	where $(\cdots)$ denotes UV finite terms. 
	\newline	
	
	Now recall that the holographic dual of a TFD state on $\M_{1} = R\times H^{d-1}$ is a two-sided AdS black hole with a horizon that has a hyperbolic geometry \cite{TFD}. The temperature of the black hole is equal to the temperature of the dual TFD state, and the AdS length scale, $\ell_{\mbox{\tiny{AdS}}}$, is set by the radius of $H^{d-1}$ (and is therefore equal to 1 in our case). For the special case where the inverse temperature is $\beta = 2\pi$, which is what we are assuming, the metric of a hyperbolic black hole in AdS is the same as that of vacuum AdS with hyperbolic slicing. Therefore, according to the conjecture cited in Eq.~\eqref{eq-proposal}, the fidelity susceptibility of a TFD state of inverse temperature $\beta = 2\pi$ on $\M_{1} = R \times H^{d-1}$ must be proportional to the volume of a maximal slice in vacuum AdS geometry with hyperbolic slicing. The metric of vacuum AdS with hyperbolic slicing is given by
	\begin{align}
	ds^{2} \, = \, - (r^{2} - 1) dt^{2} + (r^{2} - 1)^{-1} dr^{2} + r^{2} ds^{2}_{H^{d-1}} \, . \label{eq-met-ads-h} 
	\end{align}
	As mentioned earlier, since the spacetime in question is stationary, a constant time slice in the bulk has maximal volume. The volume of such a slice is given by 
	\begin{align}
	V_{\text{max}} \, = \, V_{H^{d-1}} \, \int_{1}^{r_{\text{max}}} \, dr \, \frac{r^{d-1}}{\sqrt{r^{2}-1}} \, . \label{eq-vol-hyperbolic}
	\end{align}
	where $V_{H^{d-1}}$ is the volume of $H^{d-1}$ and we have introduced a cutoff $r = \rmax$ near the boundary of AdS. In the case of $d=5$, the divergences in this volume are given by
	\begin{align}
	V_{\text{max}} \, = \, V_{H^{d-1}} \, \left( \frac{\rmax^{4}}{4} + \frac{\rmax^{2}}{4} + \frac{3}{8}\log\rmax  \right) \,  +(\cdots)  \, , \label{eq-vol-hyperbolic-div}
	\end{align}
	where $(\cdots)$ denotes UV finite terms. Note that the relationship between the bulk cutoff $\rmax$ and the UV cutoff $\epsilon$ can be determined by solving $\rmax = r(z=\epsilon)$ where $z$ is the Fefferman-Graham radial coordinate. We perform this analysis in Appendix~(\ref{app-cutoff}). From Eq.~\eqref{eq-cutoff-maps}, we find that the bulk and CFT cutoffs are related as
	\be\label{cutoff-relation-sec2}
	\rmax = \frac{1}{2\xi \epsilon} + \frac{\xi \epsilon}{2} \, ,
	\ee
	where $\xi > 0$ is some arbitrary constant. In terms of $\epsilon$, Eq.~\eqref{eq-vol-hyperbolic-div} becomes
	\begin{align}
	V_{\text{max}} \, = \, V_{H^{d-1}} \, \left( \frac{1}{64 \xi^{4}\epsilon^{4}} + \frac{1}{8\xi^{2}\epsilon^{2}} - \frac{3}{8}\log\epsilon  \right) \,  +(\cdots)  \, , \label{eq-vol-hyperbolic-div-two}
	\end{align}
	
	We now compare Eq.~\eqref{fid-div-sec2} and Eq.~\eqref{eq-vol-hyperbolic-div-two}. Assuming Eq.~\eqref{eq-proposal} to be true (with $\ell_{\mbox{\tiny{AdS}}}=1$), we are led to the following conditions. Comparison of the log-divergences implies
	
	\be
	n_{5} \, = \, \frac{448}{225} N^{\prime}_{5}\,.
	\ee
	
	Comparison of the $O(\epsilon^{-2})$ terms implies
	
	\be
	\xi^{2} = \frac{28}{25}\,,
	\ee
	
	while comparison of the  $O(\epsilon^{-4})$ terms implies
	
	\be
	\xi^{4} = \frac{7}{12} \,.
	\ee	
	Since the latter two conditions are inconsistent, we see that the conjecture in Eq.~\eqref{eq-proposal} does not hold as an exact relationship except possibly if we add cutoff dependence to the $O(1)$ constant, $n_{5}$. 
	\newline
	
	In the next section, we present our second computation of the fidelity susceptibility and compare its divergences to those in the volume of a maximal time slice in the corresponding bulk geometry.
	
	\section{Fidelity susceptibility on $R \times S^{d-1}$} \label{sec-sphere}
	
	In this section, we consider a CFT on $\M_{2} \, = \, R\times S^{d-1}$ (where the radius of $S^{d-1}$ is set to 1) and study the fidelity susceptibility of its vacuum state under a marginal deformation. We then compare it with the volume of a maximal volume slice on $(d+1)$-dimensional global AdS spacetime.\footnote{See \cite{Bak:2017rpp} for an analysis of the fidelity susceptibility of a vacuum state of a CFT on $\M_{2} = R\times S^{d-1}$ under a relevant deformation.} The comparison of the leading-order UV divergences in these two quantities was made in \cite{MIyaji:2015mia} and it was found that the leading-order divergences are consistent with the conjecture in Eq.~\eqref{eq-proposal}. In this section, we extend this analysis and study the structure of the subleading power law divergences and the logarithmic divergences in these two quantities. 
	\newline
	
	Let us start with the CFT analysis. From Eq.~\eqref{eq-vac-fid-int}, the fidelity susceptibility of the vacuum state under a marginal deformation by an operator, $\OO(\tau,\Omega)$, is given by
	\begin{align}
	G_{\lambda\lambda}^{\text{(vac)}} \, [\M_{2}] \, = \,  \int_{-\infty}^{-\epsilon} d\tau' \int_{\epsilon}^{\infty} d\tau \, \int_{S^{d-1}} d\Omega' \sqrt{g_{S^{d-1}}(\Omega')} \, \int_{S^{d-1}} d\Omega \, \sqrt{g_{S^{d-1}}(\Omega)} \quad \Big\langle \OO(\tau',\Omega') \OO(\tau,\Omega) \Big\rangle_{\M_{2}} \, , \label{eq-def-g}
	\end{align}
	where $\Omega$ denotes angular coordinates on $S^{d-1}.$ The two-point function of a marginal operator on $\M_{2}$ (up to a normalization factor) is given by \cite{Bak:2017rpp}
	\begin{align}
	\Big\langle \OO(\tau',\Omega') \OO(\tau,\Omega) \Big\rangle_{\M_{2}} \, = \, \frac{1}{\Big(\cosh(\tau-\tau') - \Omega\cdot\Omega'\Big)^{d}} \, . \label{eq-two-pt}
	\end{align}
	Now let us focus on the integrals in Eq.~\eqref{eq-def-g}. Since the entire domain of the angular coordinates is being integrated over, we can fix the angular coordinates of the operator $\OO(\tau', \Omega')$ to be the north pole in the $(\tau, \Omega)$ coordinate chart. Using this, we simplify Eq.~\eqref{eq-def-g} to get
	\begin{align}
	G_{\lambda\lambda}^{\text{(vac)}} \, [\M_{2}] \, = \,  V_{S^{d-1}} \, V_{S^{d-2}} \, \int_{-\infty}^{-\epsilon} d\tau' \int_{\epsilon}^{\infty} d\tau \, \int_{0}^{\pi} d\theta \, \frac{\sin^{d-2}\theta}{\big(\cosh(\tau-\tau') - \cos\theta \big)^{d}} \, ,\label{eq-G-sim}
	\end{align}
	where $V_{S^{d}}$ is the volume of $S^{d}$. As in the previous section, we specialize to the case of $d=5$. In this case, Eq.~\eqref{eq-G-sim} becomes
	\begin{align}
	G_{\lambda\lambda}^{\text{(vac)}} \, [\M_{2}] \, 
	=& \, \frac{4}{3} \, V_{S^{3}} \, V_{S^{4}} \, \int_{-\infty}^{-\epsilon} d\tau' \int_{\epsilon}^{\infty} d\tau \, \frac{\cosh(\tau-\tau')}{\sinh^{6}(\tau-\tau')} \, .
	\end{align}
	
	By changing the integration variables to $u = \tau-\tau'$ and $v = \tau+\tau'$, we get 
	\begin{align}
	G_{\lambda\lambda}^{\text{(vac)}} \, [\M_{2}] \, =& \, \frac{2}{3} \, V_{S^{3}} \, V_{S^{4}} \, \int_{2\epsilon}^{\infty} du \int_{-u+2\epsilon}^{u-2\epsilon} dv \, \frac{\cosh u}{\sinh^{6}u} \, ,\\
	%
	=& \,  V_{S^{3}} \, V_{S^{4}} \, \left(\frac{1}{240 \epsilon^{4}} - \frac{1}{36 \epsilon^{2}} - \frac{1}{10} \, \log\epsilon \right) \, + (\cdots) \, ,\label{eq-fid-fin}
	\end{align}
	where $(\cdots)$ denotes terms that are UV finite. 
	\newline

	Let us now perform the holographic analysis. Recall that the holographic dual of a vacuum state of a CFT on $\M_{2} = R \times S^{d-1}$ is global AdS$_{d+1}$ where the AdS radius, $\ell_{\mbox{\tiny{AdS}}}$, is fixed by the radius of $S^{d-1}$(and hence is equal to 1 in this case) with the metric 
	\begin{align}
	ds^{2} \, = \, -\big(1+ r^{2}\big) \, dt^{2} + \big(1+ r^{2} \big)^{-1} \, dr^{2} + r^{2} d\Omega_{d-1}^{2} \, . \label{eq-met-ads-g}
	\end{align}
	As in the preceding analysis, we consider a constant time slice in this geometry. The volume of such a time slice is then given by
	\begin{align}
	V_{\text{max}} \, = \, V_{S^{d-1}} \, \int_{0}^{\rmax} \, dr \, \frac{r^{d-1}}{\sqrt{1+ r^{2}}} \, ,\label{eq-vol}
	\end{align}
	where we have introduced a cutoff, $r = \rmax$, to regulate the divergences. We once again specialize to the case of $d=5$. In this case, Eq.~\eqref{eq-vol} becomes
	\begin{align}
	V_{\text{max}} \, =& \, V_{S^{4}} \, \left( \frac{\rmax^{4}}{4} - \frac{\rmax^{2}}{4} + \frac{3}{8}\log\rmax  \right) \,  +(\cdots) \, . \label{eq-vol-fin}
	\end{align}
	As in the previous section, the relationship between the bulk cutoff, $\rmax$, and the UV cutoff, $\epsilon$, can be determined by demanding $\rmax = r(z=\epsilon)$, where $z$ is the Fefferman-Graham bulk coordinate. From Eq.~\eqref{eq-cutoff-maps}, we get
	\begin{align}
	\rmax \, = \, \frac{1}{2\xi\epsilon} - \frac{\xi\epsilon}{2} \, ,
	\end{align}
	where $\xi >0$ is an arbitrary constant. In terms of $\epsilon$, Eq.~\eqref{eq-vol-fin} becomes
	\begin{align}
	V_{\text{max}} \, =& \, V_{S^{4}} \, \left( \frac{1}{64\xi^{4}\epsilon^{4}} - \frac{1}{8\xi^{2}\epsilon^{2}} - \frac{3}{8}\log\epsilon  \right) \,  +(\cdots) \, . \label{eq-vol-fin-two}
	\end{align}
	
	Since our goal is to check if the fidelity susceptibility in Eq.~\eqref{eq-fid-fin} and the volume in Eq.~\eqref{eq-vol-fin-two} are proportional as conjectured in Eq.~\eqref{eq-proposal}, let us start by assuming that this is indeed the case. That is,
	\begin{align}
	G_{\lambda\lambda}^{\text{(vac)}} \, [\M_{2}] \, = \, n_{5} \, V_{\text{max}} \, , \label{eq-conj}
	\end{align}
	where recall that $\ell_{AdS} =1$. 
	Comparing the logarithmic divergences in Eq.~\eqref{eq-vol-fin-two} and Eq.~\eqref{eq-fid-fin}, we deduce that 
	\begin{equation}
	n_{5} \, = \, \frac{4}{15} \, V_{S^{3}} \, . \label{eq-n}
	\end{equation}

	From comparing the $O(\epsilon^{-4})$ divergence, we obtain
	\begin{align}
	\xi^{4} =1 \,,
	\end{align}
	whereas from comparing the $O(\epsilon^{-2})$, we find
	\begin{align}
	\xi^{2} = \frac{36}{30} \,.
	\end{align}
	Since these two conditions are inconsistent with each other, we are led to the same conclusion as in the previous section.
		\newline

	This completes our analysis of the comparison in divergences in fidelity susceptibility and volume.
	\newline

	\section{Conclusion} \label{sec-conc}
	The goal of this paper was to study the structure of ultraviolet divergences in fidelity susceptibility in the boundary theory and compare it with the corresponding divergences in volume of extremal time slices in the bulk. Our analysis shows that the structure of these divergences is similar in both quantities but the relationship proposed in \cite{MIyaji:2015mia} does not hold exactly except possibly if we allow the $O(1)$ constant to be cutoff dependent. Although we expect to learn about the boundary dual of the volume of a maximal bulk slice in upcoming work \cite{toappear}, our results suggest that the exact bulk dual of fidelity susceptibility remains an open question. We hope that future work informed by our analysis will shed more light on the broader question of how bulk geometrical quantities and boundary information theoretic quantities are related in the context of the AdS-CFT correspondence.
	
	\section*{Acknowledgments}
	We are grateful to Thomas Hartman and Masamichi Miyaji for commenting on a draft of this paper and to Pratik Rath for useful discussions. The work of MM was supported in part by the Berkeley Center for Theoretical Physics, by the National Science Foundation (award numbers 1521446 and 1316783), by FQXi, and by the US Department of Energy under Contract DE-AC02-05CH11231 and DE-SC0014123. IS acknowledges support from the National Science Foundation (award number PHY-1707800) for part of the duration of this project.
	\appendix
	
	\section{Derivation of fidelity susceptibility} \label{app-der}
	We consider a CFT which can be analytically continued to a $d$-dimensional Euclidean space of the form $\mathcal{M} = R \times \Sigma$, where $R$ is the Euclidean time direction. We denote the Euclidean time coordinate by $\tau$ and the coordinates on $\Sigma$ by $x$. The wave-functional of a vacuum state of this theory can be represented by a path integral over the \textit{lower-half} of $\M$: $\M_{-} = R_{-} \times \Sigma$ with a boundary condition at $\tau = 0$. That is,
	\be
	\Psi_{0}[\phi] \equiv \braket{\phi}{\Omega_{0}} \, = \, \frac{1}{\sqrt{Z_{0}}} \int_{\mathcal{M}_{-}}^{\Phi(\tau=0,x)=\phi(x)} D \Phi \,\, \exp\left\{-\int_{-\infty}^{0} d\tau \, \is \, \Lg_{0}[\Phi]\right\} \, , \label{eq-vac}
	\ee
	where $h_{ab}$ is the metric on $\Sigma$, $\Lg_{0}$ and $Z_{0}$ are the Lagrangian density and the partition function of this theory respectively, and $h(x)$ denotes the determinant of the metric on $\Sigma$. The adjoint of this wave-functional is given by an analogous path integral over the \textit{upper-half} of $\M$: $\M_{+} = R_{+} \times \Sigma$ with a boundary condition at $\tau = 0$. Let us now deform our CFT by a marginal operator, $\OO(\tau,x)$. The Lagrangian density of the deformed theory is given by
	\be
	\Lg_{\lambda} \, = \, \Lg_{0} + \lambda \, \OO(\tau,x)\,,
	\ee
	where $\lambda$ is the coupling constant that is assumed to be small compared to 1. The vacuum wave-functional of the deformed theory is given by a path integral analogous to Eq.~\eqref{eq-vac} but $\Lg_{0}$ and $Z_{0}$ replaced with $\Lg_{\lambda}$ and $Z_{\lambda}$ respectively. 
	This means that overlap between the vacuum of the original theory, $\ket{\Omega_{0}}$, and that of the deformed theory, $\ket{\Omega_{\lambda}}$, is
	\begin{align}
	\braket{\Omega_{0}}{\Omega_{\lambda}} \, = \, \frac{1}{\sqrt{Z_{0}Z_{\lambda}}} \int_{\mathcal{M}} D{\Phi} \,\, \exp\left\{ -\is \Big( \int_{-\infty}^{0} d\tau \, \Lg_{\lambda}[\Phi] + \int^{\infty}_{0} d\tau  \, \Lg_{0}[\Phi]\Big)\right\} \, .
	\end{align}
	By expanding this to second order in $\lambda$ and by comparing with Eq.~\eqref{eq-def-fid-pure}, we find that the vacuum fidelity susceptibility is given by
	\begin{align}
	G^{(\text{vac})}_{\lambda\lambda} [\M] \, = \, \int_{\epsilon}^{\infty} d\tau \int_{-\infty}^{-\epsilon} d\tau^{\prime} \int_{\Sigma} d^{d-1}x \sqrt{h(x)} \int_{\Sigma} d^{d-1} x^{\prime}  \sqrt{h(x^{\prime})} \,\, \langle \OO(\tau,x) \OO(\tau^{\prime}, x^{\prime})\rangle_{\M} \, , \label{eq-vac-fid-int}
	\end{align}
	where $\langle ... \rangle_{\M}$ is the vacuum expectation value in the original CFT on $\M$. {Note that the two-point function in Eq.~\eqref{eq-vac-fid-int} diverges as the two operators approach each other, as can be seen from Eq.~\eqref{2pt}. This gives rise to the UV divergences in the fidelity susceptibility. To regulate these divergences, we have introduced a UV cutoff at $\tau = - \tau^{\prime} = \epsilon$. }
	\newline

	
	%
	
	Now let us take the product of two copies of a CFT on $\M = R \times \Sigma$ and consider a thermofield double (TFD) state of inverse temperature $\beta$ in this doubled system. This state is defined as 
	\begin{align}
	\ket{\text{TFD}_{0}} \, \equiv \, \frac{1}{\sqrt{\widetilde{Z}_{0}}} \, \sum_{n} \, e^{-\beta E_{n}/2} \, \ket{n}_{1}\ket{n}_{2} \, , \label{eq-def-tfd}
	\end{align}
	where $E_{n}$ and $\ket{n}$ are the energy eigenvalues and the energy eigenstates of the Hamiltonian, $H_{0}$, of the CFT. Moreover, the normalization constant $\widetilde{Z}_{0}$ is the thermal partition function of the CFT on $\M = R \times \Sigma$. From Eq.~\eqref{eq-def-tfd}, we note that the wave-functional of the TFD state is given by \footnote{We follow the notation used in \cite{Tom}.}
	\begin{align}
	\mbox{}_{\mbox{\tiny{1}}}\bra{\phi_{1}} \mbox{}_{\mbox{\tiny{2}}}\braket{\phi_{2}}{\text{TFD}_{0}} \, = \, \frac{1}{\sqrt{\widetilde{Z}_{0}}} \, \bra{\phi_{1}}e^{-{\beta}H_{0}/2} \ket{\phi_{2}^{*}} \, .  
	\end{align}
	
	Now let us deform the CFT by a marginal operator and consider the TFD state on two copies of the deformed theory. The overlap between the TFD state of two copies of the original theory, $\ket{\text{TFD}_{0}}$, and that of two copies of the deformed theory, $\ket{\text{TFD}_{\lambda}}$, is given by
	\begin{align}
	\braket{\text{TFD}_{0}}{\text{TFD}_{\lambda}} \, = \, \frac{1}{\sqrt{\widetilde{Z}_{0}\widetilde{Z}_{\lambda}}} \, \int D\phi_{1}\int D\phi_{2} \, \bra{\phi_{2}^{*}} e^{-\frac{\beta}{2}H_{0}} \ket{\phi_{1}}\bra{\phi_{1}}e^{-\frac{\beta}{2}H_{\lambda}} \ket{\phi_{2}^{*}} \, , \label{eq-tfd-overlap}
	\end{align}
	where $H_{\lambda}$ and $\widetilde{Z}_{\lambda}$ are the Hamiltonian and thermal partition function of the deformed theory. We can write the right hand side of Eq.~\eqref{eq-tfd-overlap} as the following path integral on $\MM = S^{1}\times \Sigma$, where the period of $S^{1}$ is $\beta$. 
	
	\begin{align}
	\braket{\text{TFD}_{0}}{\text{TFD}_{\lambda}} \, = \, \frac{1}{\sqrt{\widetilde{Z}_{0}\widetilde{Z}_{\lambda}}} \int_{\MM} D{\Phi} \,\, \exp\left\{ -\is \Big( \int_{-\beta/2}^{0} d\theta \, \Lg_{\lambda}[\Phi] + \int^{\beta/2}_{0} d\theta  \, \Lg_{0}[\Phi]\Big)\right\} \, ,
	\end{align}
	where $\theta$ is the coordinate on $S^{1}$. Expanding this to second order in $\lambda$, we deduce that the fidelity susceptibility of the TFD state under a marginal deformation  by \footnote{The same formula was also derived in \cite{Bak} for thermal states by defining the fidelity susceptibility of thermal states as the $\alpha \to 1/2$ limit of relative Renyi entropy which is defined below Eq.~(4) of \cite{Fidgen}. Note, however, that this definition is different from the one given in \cite{Uhl} where fidelity susceptibility for mixed states is defined as the $\alpha \rightarrow 1/2$ limit of ``sandwiched" relative Renyi entropy given in Eq.~(4) of \cite{Fidgen}.
		

	}
	\begin{align}
	G^{(\text{TFD})}_{\lambda\lambda} [\M] \, = \, \int_{\epsilon}^{\beta/2-\epsilon} d\theta \int_{-\beta/2+\epsilon}^{-\epsilon} d\theta^{\prime} \int_{\Sigma} d^{d-1}x \sqrt{h(x)} \int_{\Sigma} d^{d-1} x^{\prime}  \sqrt{h(x^{\prime})} \,\, \langle \OO(\theta,x) \OO(\theta^{\prime}, x^{\prime})\rangle_{\MM} \, . \label{eq-ther-fid-int}
	\end{align}
	Like the integrals in Eq.~\eqref{eq-vac-fid-int}, the integrals in Eq.~\eqref{eq-ther-fid-int} also diverge when the two operators approach each other. To regulate these divergences we need to introduce a cutoff at $\theta = -\theta' = \epsilon$ and at $\theta = -\theta' = \beta/2 -\epsilon$. 
	\newline
	
	This completes our review of the derivations of fidelity susceptibility. 
	
	\section{Relation between the bulk and boundary cutoffs} \label{app-cutoff}
	
	In Sec.~(\ref{sec-hyperbolic}) and Sec.~(\ref{sec-sphere}), we computed the UV divergences in the volume of a maximal slice in a two-sided AdS-hyperbolic black hole and in a global AdS spacetime respectively. The divergences in these quantities are due to the fact that the metric of the asymptotically locally AdS spacetime diverges as we approach the asymptotic boundary of the spacetime. These divergences are regulated by introducing a cutoff surface near the asymptotic boundary at $r=\rmax$, where $r$ is the radial bulk coordinate.  
	\newline
	
	In order to consistently compare the divergences in the fidelity susceptibility of the CFT state and in the volume of a maximal bulk slice, we need to find a relationship between the bulk cutoff, $\rmax$, and the CFT UV cutoff, $\epsilon$. According to the standard AdS-CFT dictionary, the CFT cutoff, $\epsilon$, is taken to be the cutoff value of the Fefferman-Graham (FG) radial coordinate, $z$ \cite{Fukuma}. This choice can be motivated by arguing that under scaling transformations in the boundary, the FG radial coordinate transforms as $z \to \mu z$, which is how the cutoff, $\epsilon$, transforms under scaling. In the FG coordinates, the metric of an asymptotically locally AdS spacetime has the form
	\begin{equation}
	ds^{2} \, = \, \frac{1}{z^{2}} \, \Big( dz^{2} + g_{ij}(x,z)dx^{i}dx^{j}\Big) \, . \label{eq-met-FG}
	\end{equation}
	The asymptotic boundary is situated at $z=0$ in these coordinates, and the cutoff surface near the boundary is at $z= \epsilon$. Therefore, to find the relationship between the bulk cutoff, $\rmax$, and the CFT cutoff, $\epsilon$, we need to find the relation between the bulk radial coordinate, $r$, and FG radial coordinate, $z$. To do this, we write the metrics in Eq.~\eqref{eq-met-ads-h} and Eq.~\eqref{eq-met-ads-g} in the form of Eq.~\eqref{eq-met-FG}. This can be achieved by solving
	\begin{eqnarray}
	\left( \frac{dr}{\sqrt{r^{2} \mp 1}} \right)^{2} \, = \, \left(\frac{dz}{z}\right)^{2} \, , \label{eq-rz-eq}
	\end{eqnarray}
	with the boundary condition that $r \to \infty $ as $z \to 0$. Note that $-$ sign corresponds to the metric in Eq.~\eqref{eq-met-ads-h} whereas $+$ sign corresponds to the metric in Eq.~\eqref{eq-met-ads-g}. Solving this equation yields
	\begin{align}
	r \, = \, \frac{1}{2} \left( \frac{1}{\xi z} \pm \xi z \right) \, , 
	\end{align}
	where $\xi > 0$ is an arbitrary constant of integration. This implies that the position of the cutoff surface is given by
	\begin{equation}
	\rmax \, = \, \frac{1}{2} \left( \frac{1}{\xi \epsilon} \pm \xi \epsilon \right) \, . \label{eq-cutoff-maps}
	\end{equation}
	
	
	\bibliographystyle{JHEP}
	\bibliography{Fidelity}
\end{document}